# ABOUT THE NEURONAL MECHANISM OF LATERAL HYPOTHALAMIC SELF-STIMULATION RESPONSE

**Sergey E. Murik**

Department of Physiology and Psychophysiology,  Irkutsk State University, K.Marx, 1, Irkutsk, 664003, Russian Federation (sergey_murik@yahoo.com)

### ABSTRACT

The experimental part of this study has shown that hunger motivation may be evoked by a long-term (10-180 s) continuous electrical stimulation of the "hunger center" at a current of 133.6±8.1 µA. Positive emotions were caused by electrostimulation at the same current intensity but short-term duration (0.3-0.5 s).

A positive feeling elicited by electrostimulation of the motivation center can be explained in terms of the adaptation (polarization) theory of motivation and emotion (Murik, 2001, 2005).

### INTRODUCTION

Self-stimulation response has long been considered as a model of positive emotions (Olds and Milner, 1954; Oniani, 1972; Simonov, 1972). The existence of brain structures, the electrostimulation of which gives rise to a positive subjective feeling, provided the idea of  a brain reward system (Bovard, 1962; Olds, 1958; Olds and Olds, 1965; Szabó and Milner, 1972). However, researchers have so far failed to gain a complete understanding of the mechanism for elaborating positive feelings during electrostimulation of brain cells (Barbano and Cador, 2007; Berridge, 1996; Deutsch and Howarth, 1963; Salamone et al., 2005; Wise, 2008). The brain structures assisting in self-stimulation response include also the lateral hypothalamic center (LHC) (Hunt and Atrens, 1992; Olds and Olds, 1965; Porrino et al., 1983; Szabó and Milner, 1972). Nevertheless, there is much evidence suggesting that the LHC may be regarded as the center for feeding motivation, i.e. "hunger center" (Anand and Brobeck, 1951; Mittleman and Valenstein, 1984; Wyrwicka and Doty, 1966). Hence, the same hypothalamic structures may conditionally produce both self-stimulation response and hunger motivation that seem to be associated with negative subjective feeling in the absence of food. The question of mechanism maintaining self-stimulation hunger is still an open one (Hoebel, 1979).

The viewpoint relating the autostimulation phenomenon to activation of structures relevant to the brain reward system is still under discussion (Bozarth, 1994 ; Esch and Stefano, 2004; Hoebel, 1982 ; Hoebel, 1979; Milner, 1991; Olds, 1973; Olds and Olds, 1965; Oniani, 1972; Simonov, 1972). First, there is still no evidence for the existence of structures the stimulation of which would provide an unambiguously positive response. The character of animal movements towards a response-eliciting situation often depends on certain additional factors, such as stimulation intensity, stimulation duration, functional status of the animal organism, and others (Lakomkin and Myagkov, 1980; Milner, 1973). In line with test conditions, the same brain structure has been found to respond both in a positive and in a negative way. Second, this set of factors is unlikely to be regarded as a system, because self-stimulation response may be produced in any of the brain structure. Third, even though such a reward system does exist, no explanation has yet been found for its positive emotional mechanism, so that it is beyond reason to hope for any type-specific subjective stimulation and rewarding effect in the context of any neuronal group.

More than fifty years since the discovery of the self-stimulation phenomenon, it remains unclear as to why emotionally positive responses arise from stimulation of motivatiogenic structures, e.g. hunger center. It is unlikely that hunger can be regarded as positive feeling. This raises the question as to why animals nonetheless self-stimulate the hunger center. The hunger center, in the view of those who support the ideas of cognitive psychology and systematic methodology, should be assigned to the brain reward system, though in practice the positioning of any structural-functional element of the brain has not yet advanced one jot our understanding of the psychic nature (Esch and Stefano, 2004). Just one positioning of the reward system with unexplained reward mechanism is inadequate to find these theories substantiated. The new methods for understanding the neuronal mechanism of motivation and emotion regulation (Murik, 2001; Murik, 2005; Murik, 2006) may thus provide some other explanations for this phenomenon. The author's views expressed herein are to be considered as advocating the polarization (adaptation) theory of motivation and emotion, thereby implying that electrostimulation of neural tissue of the hunger center produces changes in the metabolic (adaptation) state of the neuronal systems associated with estimation of nutrient levels. In the event that electrostimulation enhances the resistance of brain neurons, it causes positive subjective feeling and tendency (motivation) to maintain this state. We believe that these processes are just the ones providing the basis for self-stimulation response.

*The purpose of this work* has been to analyze the characteristic properties of motivational-emotional response of the rats under different parameters and modes of electrostimulation of the lateral hypothalamic area (hunger center), and to consider neuronal mechanism of self-stimulation response of the LHC in the context of the polarization theory of motivation and emotion (Murik, 2005; Murik, 2006).

**METHODS**

The work has been performed on 13 white outbred rats weighting 180-200 g. Animals were kept and used in accordance with the principles of the European Community (86/609/EEC) and the regulations of the Eastern-Siberian Scientific Center of Siberian Branch of the Russian Academy of Medical Sciences.

Two-three days prior to the electrostimulation experiments, the rats, anesthetized with calypsol (50 mg/kg), underwent bilateral stereotaxic implantation of electrodes in the LHC. The electrodes were located according to the stereotaxic atlas (Pellegrino et al., 1979). The electrodes were made of nickel-chromium or silver lacquered wire 0.25 mm in diameter. The size of the active part of the electrode was 0.44 mm$^2$. The electrodes were fastened to the skull surface with quick-hardening plastic material (acryloxide). The indifferent electrode was located in the bones of the frontal sinuses.

The electrostimulation procedure involved the application of pulsed direct current with 100 Hz frequency and square-wave pulse duration 1 ms. It was different polarity current of 2-400 μA. There were two modes of imposed electrostimulation, discontinuous and continuous. During the discontinuous mode, electrostimulation duration was no longer than 0.3-2 s, and then there was a pause lasting from 0.5 to 10-15 seconds. During the continuous mode, the duration of imposed stimulation ranged from 10 s to 3 min.

The electrostimulation experiments have been performed using the plexiglass chamber box 35×25×50 cm. Pressing the pedal on the front wall started brain electrostimulation producing self-stimulation. A rat was placed in the middle of the chamber and could move freely around the box. The rat experiments were performed no more often than once a day.

During brain stimulation, an observation was made on the behavior of the rats. Beside the pedal, in the chamber there were some other things: a piece of chalk, a piece of wood, and a plastic cork. Water and food were also placed there on a plate. The self-recording device recorded each time the animal pressed the pedal and the imposed electrostimulation periods on the tape moving at a rate of 0.5 mm/s. The experiment has been performed with satiated animals.

***Experimental procedure***: stimulation was initially imposed in a discontinuous mode at a current of 2-10 μA. Motivational-emotional responses could be judged from the behavioral symptoms: orienting-explorative response such that the animal started moving around the chamber or remained sitting there making side-to-side head motions showing searching behavior, feeding or drinking behavior, and the appearance of self-stimulation response. Then the stimulation was imposed in a continuous mode lasting from 10 s to 3 min. Continuous electrostimulation has been performed at the same current intensity that caused motivational-emotional responses in discontinuous stimulation mode.

The experiment completion was followed by checking the stimulating electrode-tip localization. Statistical processing was performed using MS EXCEL. The results are presented as M±s, wherein M is an arithmetic mean, and s is an error in mean. The significance testing was done using parametric Student's *t*-test.

**RESULTS**

Short-duration DC cathode-imposed stimulation of the LHC caused an orienting response at a current of 10-80 μA. The mean current in the whole sample was therewith 34.3±2.3 μA. The completion of electrostimulation procedure of this kind was often associated with some residual effects, such as short-lived (typically lasting 1 min) side-to-side head motions, motions doing search around the chamber, or activation of grooming behavior. Feeding or drinking responses at the indicated current intensity were activated very occasionally during and after induced discontinuous short-term stimulation, and self-stimulation response failed to be caused.

Induced continuous stimulation at current intensities similar to those indicated above often gave rise to activation of feeding behavior: a satiated rat started feeding migration around the chamber, came up to the feedbox, and began eating. The food intake process was sometimes not interrupted after current switch-off and could still go on during a short-run period.

Current intensities 2.5-3 times higher than those causing initial behavioral responses did not make many changes in the character of the animal's response to electrostimulation. For the most part, induced discontinuous stimulation caused orienting-searching responses sometimes resulting in food intake, whereas induced continuous stimulation was very often associated with feeding response activation.

Localization of stimulating electrodes in the LHC was also able to cause self-stimulation response. The minimum current intensity causing self-stimulation response in the rats was 40 μA. On average, the rats showed self-stimulation response at a current of 133.6±8.1 μA. Self-stimulation frequency was ranging from 40 to 120 presses on the pedal per minute. Induced continuous electrostimulation of the LHC, started up concurrently with self-stimulation response at a current intensity as high as that for self-stimulation, interrupted self-stimulation response in rats and activated an orienting-searching behavior resulting in feeding or drinking response (Fig. 1).

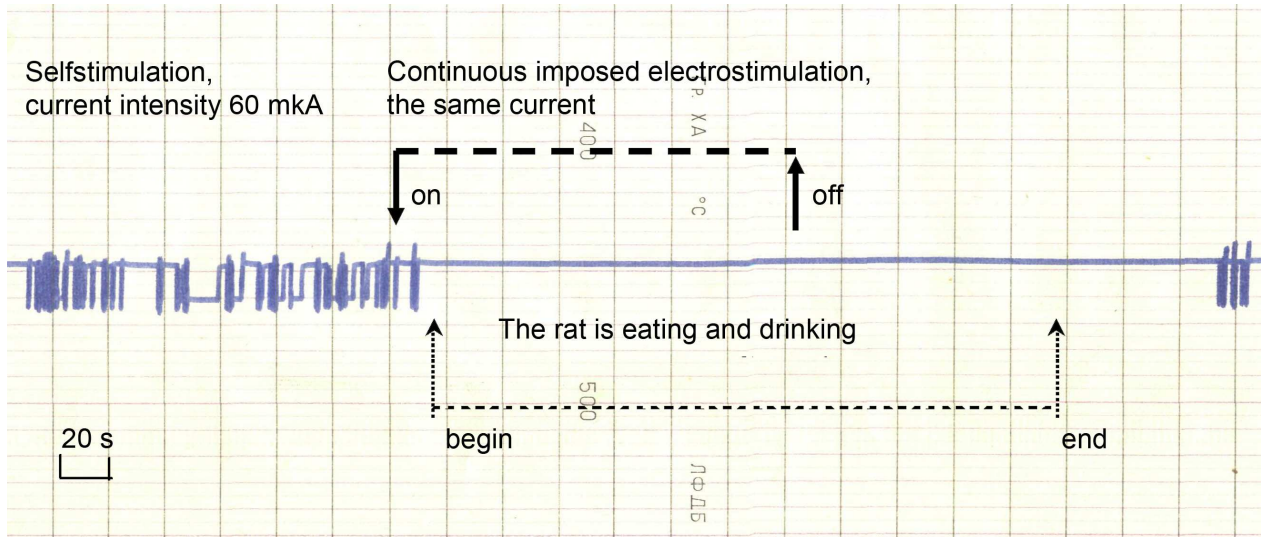

Figure 1. Mechanogram of pedal-pressing movements made by the rat during self-stimulation and continuous electrical stimulation (pulsed DC). Thick downward and upward arrows show respectively on-off switching of continuous electrostimulation of the LHC, and a thick dashed line shows duration of imposed stimulation. Thin arrows indicate the moments of activation and cessation of eating and drinking, and a thin dashed line indicates the periods during which the rat is eating and drinking.

Feeding response, as a rule, was not stopped immediately after ceasing induced continuous electrostimulation procedure. Sometime after the food intake interruption, the rats often went up to the pedal and started self-stimulation again. Repeated restarts of induced continuous electrostimulation that occurred simultaneously with self-stimulation did always stop self-stimulation and activated feeding behavior though the rats could still press the pedal mechanically some of the time (Fig. 2).

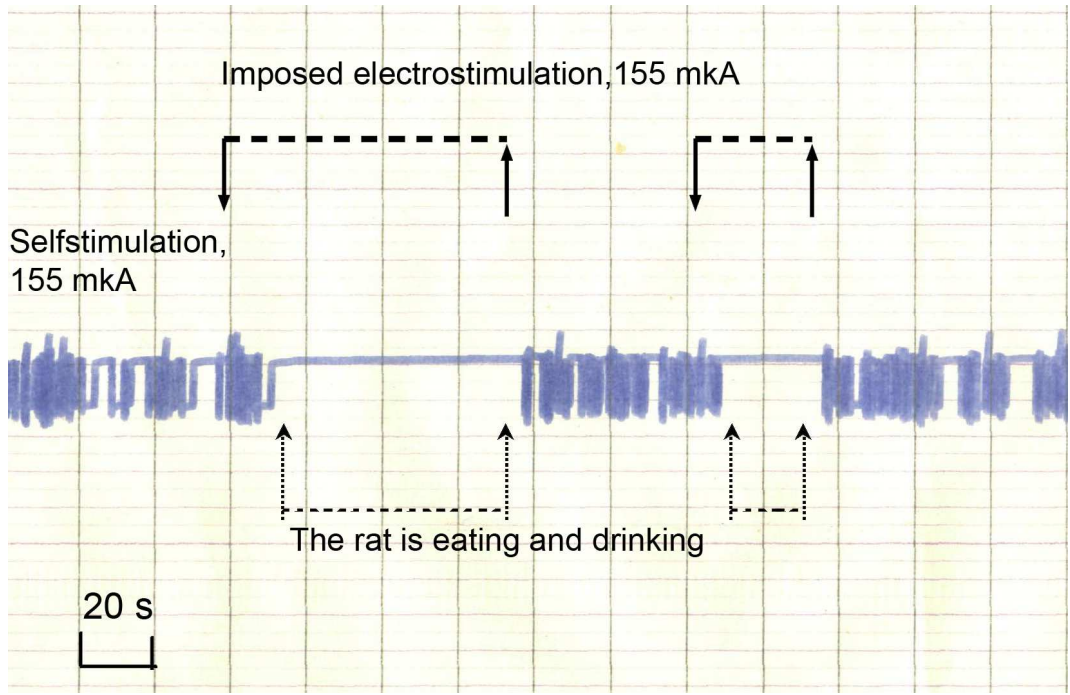

Figure 2. Mechanogram of pedal-pressing movements made by the rat during self-stimulation and continuous electrical stimulation (pulsed DC). The notations are the same as in Fig. 1.

Analysis of rat behavior during anode electrostimulation of the LHC under pulsed DC power showed that the nature of electrostimulation response was similar to that generated by the cathode and resulted in activation of similar orienting-searching responses. However, there was some dissimilarity concerning the fact that neither continuous nor discontinuous stimulation with low intensity current ever caused feeding behavioral response in satiated animals. Some feeding responses were occasionally observed in animals after the stimulation imposed at a high-intensity current of 150-400 µA. In rare instances, the anode current could elicit steady self-stimulation that was characterized by infrequent pressing of the pedal. The change in the current flow from (-) to (+) during cathode self-stimulation tended to stop the autostimulation (Fig. 3). An animal walked away from the pedal and started to move around the chamber, sometimes reaching the pedal again and performing single pressing motions. Self-stimulation response recommenced each time such motion could provide cathode current to the stimulating electrode.

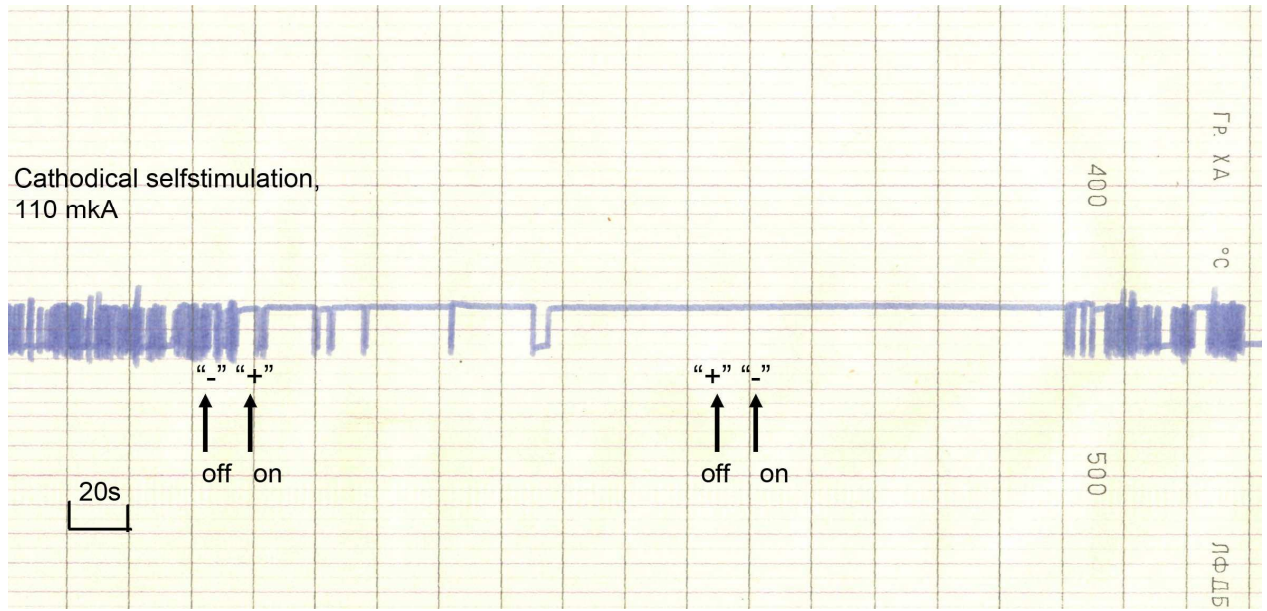

Figure 3. Mechanogram of pedal-pressing movements made by the rat during self-stimulation and under change of current polarity. The arrows show on-off switching of cathode («-») and anode («+») stimulation.

Hence, imposed electrostimulation of the LHC activated behavioral patterns of motivation-emotional responses in rats. Low-intensity current caused activation of both orienting-searching and feeding responses, though the latter were few and only became more pronounced during induced continuous electrostimulation. In a current of invariable intensity, electrostimulation of the LHC could cause not only activation of feeding behavior but also self-stimulation response. A prerequisite to the change from one pattern of the animal's motivational-emotional behavior to another is the character of pulse stimulation. Electrostimulation by a relatively short (0.3-0.5 s duration) series of pulses may promote self-stimulation responses whereas continuous electrostimulation at a 100 Hz pulse current frequency activates foraging behavior in combination with drinking. Most of these effects were generated by the action of a current of negative polarity.

**DISCUSSION**

On the evidence provided by J.P. Huston (Huston, 1971; Huston, 1972), the lateral hypothalamic stimulation with long trains of pulses provides greater milk intake than that with short trains, though the rats prefer the latter type of stimulation. In other words, the feeding behavioral response was activated by longer stimulation than the self-stimulation response. With the same electrode positions in the lateral hypothalamus, threshold of activation in motivational brain structures, considered in terms of influence on feeding and drinking processes, appeared to be different from that of positive emotional support – self-stimulation. This is sometimes interpreted as evidence that motivations have their own morphophysiological substrate existing separately from emotions (Ball, 1969; Olds and Olds, 1965; Simonov, 1972; Vanshtein and Simonov, 1979).

The experiments performed by N.G. Mikhailova and K.Yu. Sarkisova (Mikhailova and Sarkisova, 1977) showed the same trend in behavioral response during both hypothalamic polarization by gradually increasing direct current and stimulation by rhythmic current of increasing intensity when weak stimulation caused generalized searching activity without addressing any of goal objects in the chamber: food, water, animal of the opposite sex, etc. These extrinsic stimuli only become effective with increasing the stimulation intensity that caused an animal to start eating, sometimes drinking, gnawing etc. Further increase in rhythmic or direct current intensity caused self-stimulation response. In the context of these studies, the threshold of positive emotional support appeared to be higher than that of purposeful emotionally motivated behavior.

Our experiments have shown that the same pulse current intensity may cause both self-stimulation response and feeding motivation. The change from one behavior pattern to another depended on the character of electrostimulation. The short square wave pulse trains provided self-stimulation response, and the long ones involving continuous stimulation of the LHC activated feeding behavior. This is consistent with the results obtained by J.P. Huston (Huston, 1971; Huston, 1972).

If the lateral hypothalamus and other structures producing self-stimulation are related to the reward system (Bovard, 1962; Olds and Olds, 1965; Szabó and Milner, 1972), then the animals would be apt to seek continuous stimulation; instead, the animals avoid continuous stimulation of "electropositive points". A.I. Lakomkin and I.F. Myagkov (Lakomkin and Myagkov, 1980) reported that when the self-stimulation conditions were changed so that an animal would be able to control both frequency and duration of brain stimulation, it would also be able to achieve brain rhythmic stimulation, i.e. an animal would press the pedal, keep its pressed for some time, unpress, and then press the pedal again. Besides, long-term stimulation of the emotionally positive point of the hypothalamus brain (prolonged press of the pedal) often causes negatives responses, and the longer-term the stimulation is, the longer is the interval at which an animal presses the pedal.

Our research has shown that long-term stimulation of the LHC as a positive point of the hypothalamus produces motivational state like hunger thus providing a completely reliable basis for attributing this brain area to the "hunger center" (Anand and Brobeck, 1951; Mittleman and Valenstein, 1984; Wyrwicka and Doty, 1966). The fact that any motivation is associated with negative emotional feeling enables one to understand what makes animals avoid long-term electrostimulation of the LHC. However, it gives rise to another question as to why though they do it with short pulse trains. It is unlikely that relatively short-term stimulation activates some brain reward systems, because these systems may be either well activated during longer-term stimulation whose electric current intensity furthermore remains the same. Even if it is granted that the LHC is a motivatiogenic structure and the LHC stimulation only activates motivated state, there appears to be no explanation as to why animals tend to cause hunger motivation that should produce negative rather than positive egocentric emotions.

The phenomenon of stimulation of the "hunger center" may be explained in the context of polarization theory of motivation and emotion, which is advocated by the author (Murik, 2001; Murik, 2003; Murik, 2005; Murik, 2006). Analysis of the literature data, particularly of those obtained by the St.-Petersburg physiological school of N.E.Vvedenskii–A.A.Ukhtomskii–L.L.Vasilyev, suggests that the processes associated with nervous tissue response to various stimuli have generally similar pattern. These processes may also be considered as consecutive steps of the general process of neuronal adaptation showing regular trends in the degree of cellular membrane polarization (Fig. 4) around which the *polarization (adaptation) theory* of motivation and emotion has been developed.

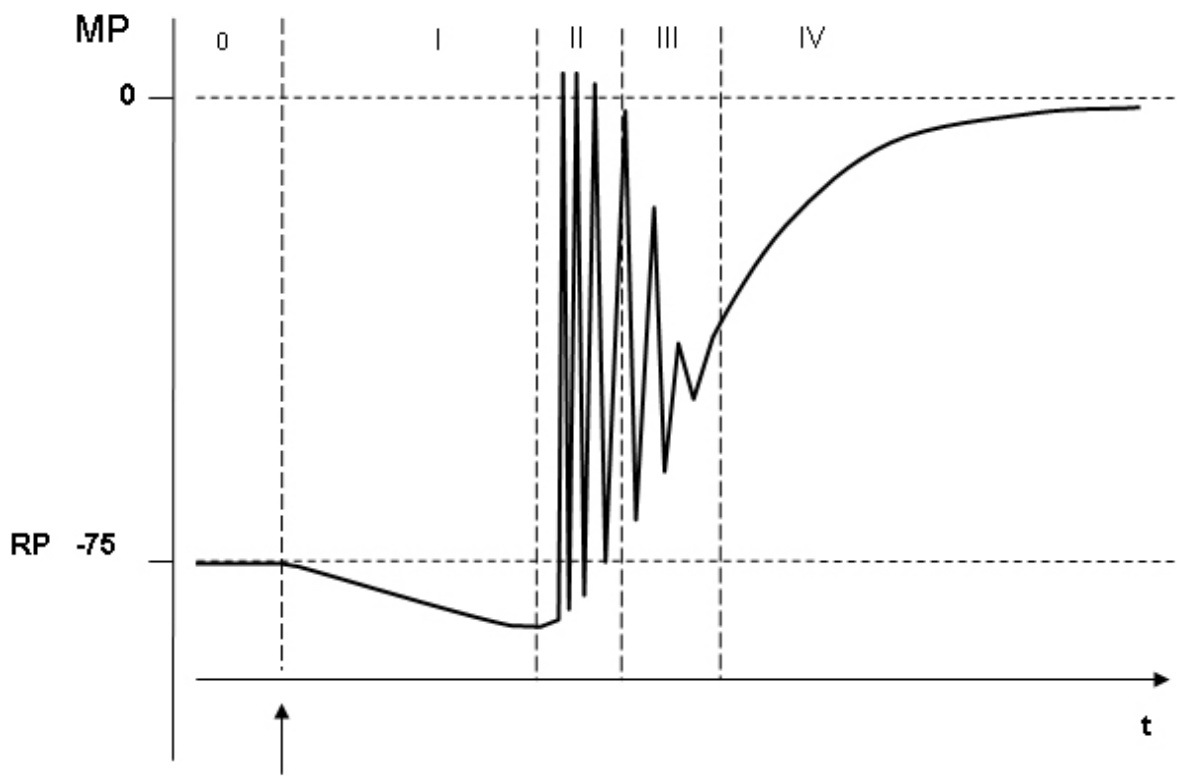

Figure 4. Variation in membrane potential (MP) and change in the state of neural tissue under the action of stimuli with time (t). The state goes through the following stages of change: 0 – rest, I – hyperpolarization silence ("inhibition"), II – posthyperpolarization excitation, III –

depolarization excitation, IV – depolarization suppression ("inhibition"). An arrow indicates the beginning of action of the stress-stimulus. RP – resting potential level.

Most likely, the beginning of stress-factor action on a neuronal cell mobilizes its endocellular adaptation reserves that enhances cell resistance to unfavorable factors and improves its functional capabilities (lability,(Vvedenskii, 1901)). In terms of electrographic activity, it implies the appearance of hyperpolarization wave (Fig. 4, I). N.A. Vvedenskii(Vvedenskii, 1901) reported it as a prodromic or electropositive step, and L.L. Vasilyev(Vasilyev, 1925) – as an antiparabiotic. There is some evidence for increase during *hyperpolarizing inhibition* of neural resistance (Murik, 2005; Murik, 2006).

Inefficiency of endocelluar mechanisms of adaptation causes the neurons to start generation of nervous impulses by mechanism of *afterhyperpolarization response*(Skrebitskii, 1977) or *post-anodal exaltation (anode-break excitation)*(Andersen and Sears, 1964). The neurons generate nervous impulses and in doing so they mobilize neuronal chains of systems mechanism of adaptation. However, an excitation like *afterhyperpolarization response* (Fig. 4, II) may be considered as *an excitation against the background of relatively good metabolic and functional state* of neural cells (Murik, 2003; Murik, 2005; Murik, 2006).

Long-term or adverse effect on a cell gradually depletes its metabolic reserves thus leading to decrease (depolarization) in resting potential (RP) of a neuron. A sustained depolarization of RP activates cascades of endocellular pathogenetic biochemical processes, lipid peroxidation in particular, and decreases lability (functional capabilities) of a cell (Murik, 2003; Murik, 2006). Hence, an excitation against the background of sustained depolarization of RP, i.e. *depolarization excitation*, is here considered as *an excitation against the background of relatively bad metabolic and functional state* of a neuron. A stage of *depolarization excitation* in Fig. 4 is represented as III. This is precisely the type of excitation to be regarded as *motivational excitation* in the context of *polarization theory*. Such process occurs extensively in the nervous system causing its rejection by the organism that just looks like purposeful behavior. Hence, the essence of motivated behavior of the organism as a whole lies in its striving for recovery of a good metabolic and functional state of depolarized brain cells. If stress-factor is not eliminated early, the adaptation reserves will be completely depleted, and depolarization will be so extensive as to disable the generation of nervous impulses (Fig. 4, IV). Depolarization-induced suppression will be combined with the development of even more negative endocellular biochemical processes regularly giving rise to activation of apoptosis mechanism. N.E. Vvedensky reported this stage as parabiotic suppression. We find *depolarization suppression* phenomenon in a nervous system even more undesirable than depolarization excitation, because it is conceptually associated with failure of neuronal adaptation, which would most likely result in cellular necrosis.

Therefore, within the context of polarization theory, motivations and emotions are psychic phenomena arising from internal and external stimuli, reflecting changes in metabolic, and functional (adaptation) state of the brain nervous tissue. This approach implies that there is a close and intimate link between motivation and emotion substrates in the brain. Motivated states and negative emotions are predetermined by appearance of *depolarization excitation. However, the reduction of depolarization excitation* as well as *its change to posthyperpolarization excitation or hyperpolarization suppression* would be regarded as a "reward" for the nervous system and organism as a whole and underlie the appearance of positive emotions because of being responsible for enhancing adaptation reserves of nerve cells in the brain.

Electostimulation of the LHC by currents of different intensity, duration and polarity in the context of polarization theory produces phenomena wherein relative long-term continuous cathode stimulation of the LHC provides an appearance of *depolarization excitation* in the food-related brain systems (Zambrzhitskii, 1989), which is followed by negative subjective state of hunger and the tendency to subdue it by eating. From this standpoint, the LHC self-stimulation response may also provide a model of motivated state only if both actualization and hunger satisfaction are realized through on/off electrostimulation switching operations. Short-term electrostimulation (0.3-0.5 s) causes the appearance of relatively local *depolarization excitation* and thus motivated state. The current switching-off process involves repolarization of neuronal elements in the LHC-related brain system and hence reduction of motivated state and appearance of positive emotions.

Therefore, self-stimulation response in the context of polarization theory of motivation and emotion does not provide a model of purely positive emotions, because it is a pattern of complex motivated behavior the components of which are both negative and positive emotions. Short-term electrostimulation of the LHC causes something like hunger that disappears and gives rise to positive emotions like satiety after switching-off the current source. The animals press the pedal to switch current on tending more to the state after the current is halted than to the state arising during electrostimulation. Most likely positive emotions caused by apparent reward (current switch-off) are much stronger than negative emotions produced by electrostimulation. Long-term stimulation of the LHC activates the natural motivated state of hunger associated with strong negative emotions. That is why an animal starts eating, trying to avoid electrostimulation if possible (Lakomkin and Myagkov, 1980).

Since membrane potential depolarization is caused by current flowing through the cathode with an electrode located outside of the excitable tissue (Khodorov, 1969), it is the cathode current that may rapidly cause depolarization excitation, whereas the anode exerts an opposite effect causing membrane hyperpolarization. The anode current may cause excitation only under certain conditions: current switch-off mechanism related to anode exaltation (posthyperpolarization response). It is much more difficult to achieve depolarization excitation with the anode current. To do this probably requires high current intensity and some long-duration impulses.

Hence, various behavior patterns (feeding responses and self-stimulation) obtained by the LHC stimulation ranging either in intensity or character are scarcely evidence for the incontestable existence of separate brain substrates and mechanisms for demands, motivations and emotions. Most likely neural substrates of motivational and emotional states in terms of polarization (adaptation) theory provide an integral unit, related to the change of metabolic and functional (adaptation) state of the same neurons participating in stimulus perception. Positive emotions related to demand satisfaction are assumed to arise from a common mechanism of "drive reduction" (Hull, 1943 ; Miller and Dollard, 1941). *We suggest that the center of food motivation, negative feelings of hunger and positive feelings of satiety refers to the same structure related with an analyzer of the internal nutrient level. All these psychic phenomena respectively have the same neuronal substrate. The only difference between them is a qualitative temporal variation of the adaptation and thus polarization state of the neurons composing this substrate (analyzer).*

Another explanation may also be given to self-stimulation response mechanism in this context. In particular, the occurrence of inhibitory processes in the nervous system in the course of self-stimulation, associated with electrophysiological changes (Mikhailova, 1971a; Mikhailova, 1971b) suggests that relatively short-term activation of the LHC may also cause generalized processes of neuronal "hyperpolarization suppression". Although the idea that self-stimulation is a tendency for activation which causes post-synaptic inhibition had been already reported (Myslobodskii, 1970), the biological mechanism of this tendency remained unknown. As far as adaptive state is concerned, "hyperpolarization suppression" is the evolution of prime metabolic and functional state (Murik, 2003; Murik, 2006). Because of this, the enhancement of inhibitory hyperpolarization processes of the nervous system should be followed by the occurrence of positive subjective feeling. If reinforcing effects of self-stimulation zones are associated with the enhancement of hyperpolarizing processes in the brain structures, then the occurrence of motivational state in the course of continuous electrostimulation of the same points may be attributed to the brain cell transition to the next adaptive stage of posthyperpolarization excitation during a relatively long stimulation (Fig. 4, II). Under these circumstances, electrostimulation should be interrupted for neuronal reserves adaptive recovery and for the avoidance of appearance of depolarization excitation (Fig. 4, III) or depolarization suppression (Fig. 4, IV) in the brain.

Hence, it is suggested that short-train electrostimulation (0.5 s or less) enhances the adaptation reserves of neurons in the hunger center and in related brain systems whereas long-train electrostimulation depletes them. That is why the animals self-stimulate the brain with short-train pulses and avoid long-term stimulations or otherwise activate feeding behavior as a way to enhance adaptation reserves of neurons in the hunger center.

Specifying the character of polarization and adaptation shifts in the nervous system during electrostimulation of the LHC with currents of different intensity, polarity and duration invites further investigations though the already available data may well suffice to assure the plausibility of surmises.

Hence, the results of self-stimulation experiments considered in the context of polarization (adaptation) theory of motivation and emotion may provide some new ways to understand the mechanism of self-stimulation response and to explain the neurophysiological basis of purposeful behavior.

## ACKNOWLEDGMENTS

I thank V.Moliadze for technical assistance in preparation this article and O.L.Gamboa for language corrective.
## ABBREVIATIONS

Lateral hypothalamic center (LHC), direct current (DC), resting potential (RP), membrane potential (MP).